\documentclass[journal]{IEEEtran}
\usepackage{wrapfig}
\usepackage[dvipdfmx]{graphicx}
\usepackage{amsmath}
\usepackage{amssymb}
\usepackage{amsfonts}
\usepackage{comment}
\usepackage{bm}        
\usepackage{verbatim}   
\usepackage[left]{lineno}

\usepackage[dvipdfmx, bookmarks=true, colorlinks=true]{hyperref}

\newcommand{\Equref}[1]{Equation.~(\ref{#1})}
\newcommand{\Figref}[1]{Figure.~\ref{#1}}

\ifCLASSINFOpdf
\else
\fi
\begin{document}
%
\title{Performance of a large aperture GEM-like\\ gating device for the International Linear Collider}
%
%

\author{
        \IEEEmembership{T. Ogawa on behalf of the LC-TPC collaboration}
\thanks{T. Ogawa is with  the Graduate University for Advanced Studies (SOKENDAI), Tsukuba 305-0801, Japan e-mail: ogawat@post.kek.jp }
}

%
%

\markboth{2017 IEEE Nuclear Science Symposium Conference Record}
{Shell \MakeLowercase{\textit{et al.}}: Bare Demo of IEEEtran.cls for IEEE Journals}
%



\maketitle





\begin{abstract}
One of the potential problems of a Micro-Pattern Gaseous Detector (MPGD)-based Time Projection Chamber (TPC) is the Ion back Flow (IBF): ions generated through the avalanche amplification process flow back to the drift volume of the TPC and disarrange an electric field inside it. Consequently non-negligible degradation of azimuthal spatial resolution is caused due to this IBF. Meanwhile, it is necessary to collect primary ionized electrons to maintain intrinsic performance of the MPGDs.   
The MPGD based TPC is currently planned to be used as a central tracking detector of the International Large Detector (ILD), which is one of the detector concepts for the future International Linear Collider (ILC) project, and which requires fine azimuthal spatial resolution of less than 100 ${\rm \mu m}$ over the drift length of the TPC to attain high momentum resolution. 
Because of a unique beam structure of the ILC, the IBF is a critical issue for the realization of the ILD-TPC. Not only to suppress the ion back-flow to the drift volume, but also to allow the primary electrons pass through, a large aperture GEM-like gating device has been developed. 
Several bench tests for confirming the performance of the gating device have been conducted, besides that, beam test with the full detector module equipped with the gating device was carried out to verify the resolution that the full module can provide. As a result, it turned out that the developed gating device fulfills requirements for maintaining the performance of the MPGD based TPC, and it has sufficient performance for the central tracker of the ILD at the ILC.        
\end{abstract}


%
\IEEEpeerreviewmaketitle

\section{Introduction}
%
%
%
%


\IEEEPARstart{F}{or} the future International Linear Collider (ILC) project a detector concept called the International Large Detector (ILD) is proposed, where a Micro-Pattern Gaseous Detector (MPGD)-based Time Projection Chamber (TPC) is a candidate of the central tracking detector to employ several benefits of the TPC such as track pattern recognition by continuous tracking, charged particle identification and low material budget to avoid energy loss before calorimeters. Requirements of performance for the ILD-TPC are clearly given from the physics point of view under the assumption for a Minimum Ionizing Particle (MIP) \cite{ILD}, which are as follows:

\begin{itemize}
 	\item Momentum resolution with the stand-alone TPC must achieve $\sigma_{p_{T}}/p_T \sim 1 \times 10^{-4} p_T ~{\rm GeV}$.
	\item To accomplish above momentum resolution, azimuthal $r\phi$ spatial resolution must be $\sigma_{r\phi} < 100 ~{\rm \mu m}$ over the drift length of the TPC, which is given in terms of the well-known momentum resolution formula \cite{mom} under the assumption of realistic detector parameters.     
	\item Z resolution of $\sigma_{Z} < 400 \!\sim\! 1400 ~{\rm \mu m}$ is necessary for particle separation.
	\item dE/dx resolution $\sigma_{dE/dx} \sim 5\%$ is required to perform charged particle identification, . 
\end{itemize}

Especially the requirement that the $r\phi$ spatial resolution fulfill $100 ~{\rm \mu m}$ over the drift length of the TPC is possible to achieve with usage of the MPGDs. On the other hand a traditional Multi-Wire Proportional Chamber (MWPC) can not reach the required spatial resolution under the condition of a high magnetic field of $\sim$ 3.5~T because of the $\bm{E}\! \times \!\bm{B}$ effects \cite{ILD-MWPC}. 
In addition the usage of the MPGDs can provide several advantages, by itself, such as suppression of the Ion Back Flow (IBF) whose explanation will be given in the next section and no $\bm{E}\! \times \!\bm{B}$ effects.  

\section{Ion back flow and\\ requirements to a gating device}

Ions generated through the ionization process (primary ions) by charged tracks and the avalanche amplification process (secondary ions) disarrange electric field and distort observed charged tracks, which is an innately and classical problem of any TPC. 

\begin{figure}[htbp]
\begin{center}
	\includegraphics[width=84mm]{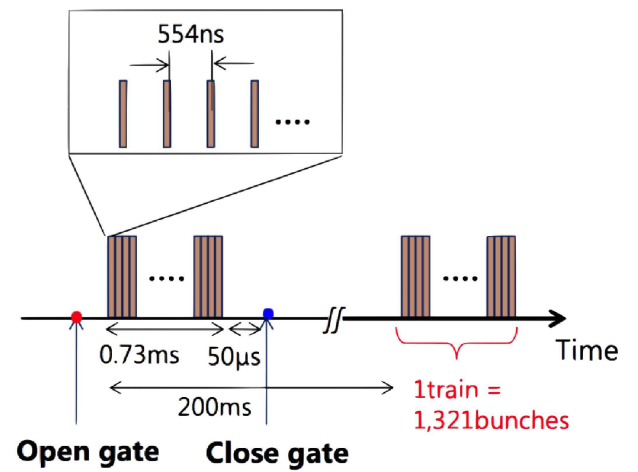}
\caption{A schematic view \cite{Seguei} of the ILC beam structure. During 1 millisecond beam-crossing duration the gate is opened to acquire the primary ionized electrons, and closed after acquisition to absorb the ions generated in the amplification devices.}
\label{fig:fig1}
\end{center}
\end{figure}

Particularly for the MPGD-based TPC the gas amplification is performed at near the endplate of the TPC. The generated ions slowly flow back to the drift volume of the TPC along the electric field. On the other hand, the ILC beam structure is unique, which is designed so that 1 millisecond beam-crossing duration is provided at every 200 ${\rm msec}$ (5~Hz), and 1321 bunches which compose 1 beam train are continuously collided during 0.73 ${\rm msec}$, which is illustrated in \Figref{fig:fig1}. 

Mobility of an isobutane ion, generally isobutane ions (${\rm iC_4H_{10}^+}$) are assumed to exist under the usage of the T2K gas (${\rm Ar : CF_4 : iC_4H_{10} = 95:3:2}$) \cite{t2k} because of ion exchange with argon ions (${\rm Ar^+}$), is 1.6 ${\rm cm^2 / (V \cdot sec)}$ \cite{isoMob} and its drift velocity is approximately 370~${\rm cm/sec}$ with the drift field of 230~${\rm V/cm}$. Therefore a thinner radial ion disc of roughly 0.3 cm thickness is assumed to be created after the collision of 1 beam train, and the three ion discs regularly float inside the volume of the TPC because the distance between the ion discs is about 75~cm and the drift length of the TPC is designed as 220~cm.           

\subsection{Requirement for ion blocking}

\begin{figure}[htbp]
\begin{center}
	\includegraphics[width=86mm]{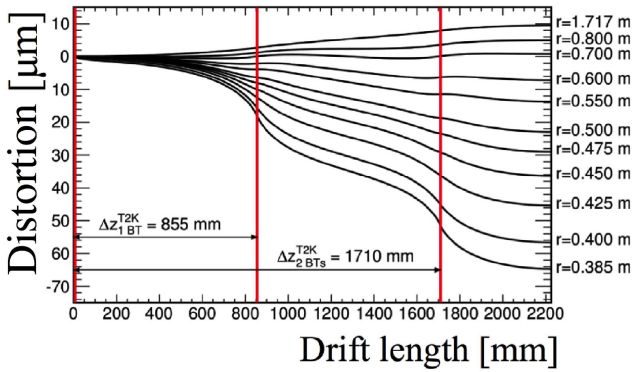}
\caption{The distortion of the $r\phi$ spatial resolution due to the three ion radial discs floating inside the TPC. At the innermost layer it reaches 60 ${\rm \mu m}$ where the IBF=1 is assumed \cite{Thorsten}, which is equivalent to the situation of the gas gain of 3,000 and the ion blocking power of $O(10^{-4})$.}
\label{fig:fig2}
\end{center}
\end{figure}

In past studies the track distortion due to the three ion discs was evaluated. \Figref{fig:fig2} shows the distortion of the track, and it turned out that the distortion will reach about 60 ${\rm \mu m}$ at the innermost layer of the ILD-TPC. The evaluation was actually given under the assumption of the IBF=1 (see a section 6.3 of \cite{Thorsten}, the detail description is given) being consistent with the condition that the gas gain of 3,000 and the ion blocking power of $O(10^{-4})$ are assumed. This $O(10^{-4})$ can be, however, achievable using a Micro-MEsh Gaseous Structure (Micromegas) \cite{Micro} or a triple- Gas Electron Multiplier (GEM) \cite{gem} structure with an optimized setting for getting the better ion blocking power \cite{MicroIBF, GEMIBF}. In the case a double-GEM structure is used, it might reach only $O(10^{-2})$. To make the distortion due to the IBF negligible the ion blocking power of $O(10^{-4})$ is at least necessary, which is one critical requirement for the gating device.      



\subsection{Requirement for electron transmission}

Meanwhile behavior of the spatial resolution established for the MPGD-based TPC have been well-understood, and the asymptotic formula describing its behavior is given as follows \cite{rphiReso}
\begin{eqnarray}
\sigma_{r\phi} (Z) \; = \; \sqrt{ \sigma_{0}^{2} + \frac{C_{d}^{2}}{N_{eff}} \cdot Z } ~~~,
\label{eq1}
\end{eqnarray}
where the second term is composed of a diffusion constant $C_d$ and $N_{eff}$ representing the effective number of electrons that is a value including the number of primary ionized electrons, ionization statistics and gain fluctuation \cite{kobaNeff}. The first term is a constant decided by a finite readout pad width and also $N_{eff}$ \cite{rphiReso,Ryo}. Thus, the overall behavior of the spatial resolution is proportional to $1/\sqrt{N_{eff}}$.

\begin{figure}[htbp]
\begin{center}
	\includegraphics[width=86mm]{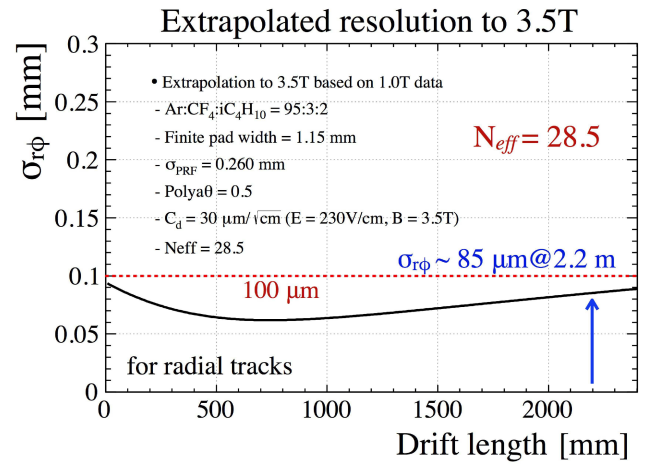}
\caption{The plot shows the extrapolated $r\phi$ spatial resolution to 3.5~T as a function of the drift length \cite{Ryo}. The evaluation was done for radial stiff tracks provided from 5~GeV electron beam, and the detector was configured with a double-GEM structure without any gating device.}
\label{fig:fig3}
\end{center}
\end{figure}

Achievable spatial resolution with the double-GEM structure without any gating device has also measured, and it confirms that the spatial resolution extrapolated to a higher magnetic field of 3.5~T is possible to reach approximately 85 ${\rm \mu m}$ as given in \Figref{fig:fig3}. Assuming the gating device reaches electron transmission rate of 80~\%, the spatial resolution can still maintain below 100~${\rm \mu m}$ and satisfy the requirement of the ILD-TPC because the electron transmission rate of 80~\% correspond to degradation of the spatial resolution by about 12~\% according to \Equref{eq1}. This electron transmission rate of 80~\% is another requirement for the gating device.

\section{A large aparture gating device}
\begin{figure}[htbp]
\begin{center}
	\includegraphics[width=84mm]{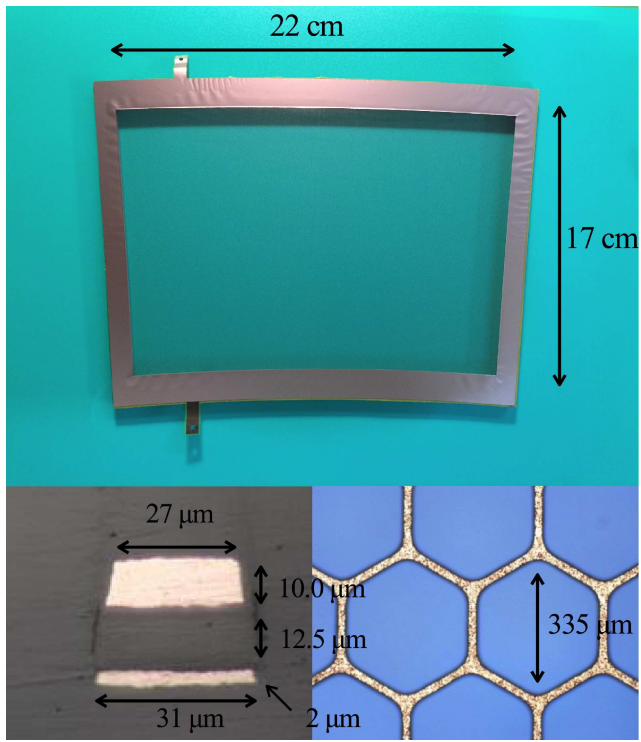}
\caption{A picture of the $22\times17cm^{2}$ large sensitive area gating device which is the real module size of the ILD-TPC. The geometric aperture is about 82.3\%, and its thickness is 25 ${\rm \mu m}$. To maximize the geometrical aperture, the hole shape of the gating device is formed as a hexagon.}
\label{fig:figGate}
\end{center}
\end{figure}

The first prototype large aperture gating device whose geometrical aperture reach 82.3~\% was manufactured \cite{Arai} cooperating with Fujikura Ltd. \cite{Fujikura} and reported at IEEE14 \cite{Ikematsu}. Afterward a 22$\times$17~${\rm cm^{2}}$ large sensitive area gating device displayed in \Figref{fig:figGate} was also produced, which corresponds to a real module size of the ILD-TPC.     

\section{Electron transmission and ion blocking}

The real measurement of the electron transmission rate using X-rays from the $^{55}{\rm Fe}$ radiation source was also reported at IEEE14. The experimental procedures and the detailed results can be found in the conference record \cite{Ikematsu}. In order to predict performance of the gating device in the higher magnetic field, simulation studies have been modified. The simulation is conducted with softwares Gmsh \cite{Gmsh} and Elmer \cite{Elmer}, which are respectively used for modeling of geometries and field calculation based on the Finite Element Method, and $\rm{Garfield^{++}}$ \cite{Garf} for electron trajectory tracking by dealing with electron transport properties of arbitrary gas mixtures, which are implemented with Magboltz \cite{Magb}. The electron trajectory tracking is performed with the Avalanche Microscopic tracking (detailed description can be found in \cite{Heinrich}) installed in the $\rm{Garfield^{++}}$, where time step calculation of the microscopic tracking is independently corrected so as to include continuous variation of electric field depending on time. 

\begin{figure}[htbp]
\begin{center}
	\includegraphics[width=84mm]{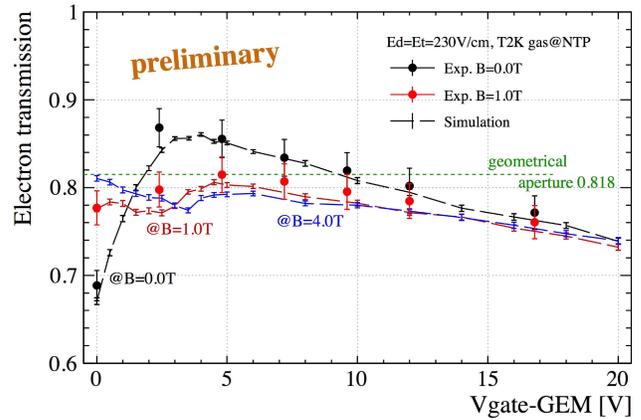}
\caption{The experimental and simulation result of the electron transmission. The points show data for 0 (black) and 1~T (red) and the lines correspond to the simulation results. The green dotted line shows the geometrical aperture of the geometry model used for the simulation.}
\label{fig:fig5}
\end{center}
\end{figure}

\subsection{Electron transmission measurement}

\Figref{fig:fig5} shows the electron transmission rate in which the experimental data and the simulations are compared in the magnetic filed of 0 and 1~T. The simulation can reproduce the experiments to some extent. The result of 4~T is also simulated to predict the performance under the higher magnetic field, and it turns out that the electron transmission rate of 80~\% could be achievable with $\Delta V \sim$  0~V operation. The reason a peak around 4~V gradually disappears under the higher magnetic field is presumably that an electron reaches a boundary of a rim of the gating device is lead to a direction of the hole center while being accelerated near the entrance of the hole because of electric force lines, and in the case a place the electron reaches is far enough from the rim, the electron is not pulled near the rim to the direction of the rim. Once the magnetic field is added, the motion of the electron perpendicular to the magnetic filed is restricted. Even though the electron is accelerated toward the hole center, the place the electron can reach is not sufficiently far from the rim. As a consequence the electron reaches the wall of the gating device by being pulled near the rim and the peak disappears.                  

\subsection{Ion blocking measurement}

\begin{figure}[htbp]
\begin{center}
	\includegraphics[width=84mm,height=57mm]{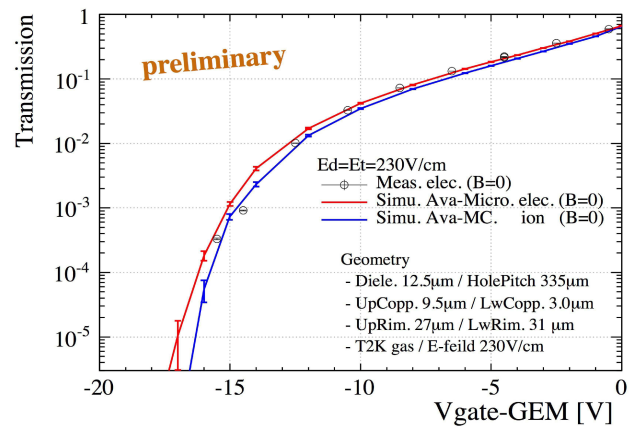}
\caption{The plot shows the result of transmission rate. The points are experimental data for the electron, and the red and the blue lines are simulation results for the electron and the ion respectively.}
\label{fig:fig6}
\end{center}
\end{figure}

\Figref{fig:fig6} shows a result of the ion blocking power evaluated by using an electron instead of an ion. A diffusion constant of an electron under the usage of the T2K gas is roughly 300~${\rm \mu m/ \sqrt{cm}}$, and one of an ion is about 150~${\rm \mu m/ \sqrt{cm}}$, where the condition of the T2K gas with the drift field of 230~V/cm is assumed. When the magnetic field is added, the diffusion constant of the electron varies dramatically depending on magnitude of the magnetic field, which is described as     
\begin{eqnarray}
\frac{D_T(\omega)}{D_T(0)} \; = \; \frac{1}{1 + (\omega\tau)^2} ~~~,
\end{eqnarray}
where $D_T$ shows the transverse diffusion constant, and $\omega$ and $\tau$ represent cyclotron frequency and mean free time defined as $\omega = (q B)/m$ and $\tau=m\bar{v} / (q E)$ ($q$, $E$, $B$ $\bar{v}$ and $m$ are a charge, a electric and magnetic field, mean velocity and mass of a particle), respectively. A remarkable thing is that the diffusion constant of the ion does not change largely even though the magnetic field is given because of its slow velocity due to massiveness, which means $\omega\tau$ of the ion is close to 0. Therefore a measurement under no magnetic field using the electron gives us a kind of the lower limit for the ion blocking power because the diffusion of the electron is twice larger than that of the ion. 


Because the field is almost closed when more than -10~V is provided to the gating device, large amount of the electron is generated with a 266~nm UV-laser system for conducting the measurement. The simulation results for the electron and the ion are also given on the plot, and the simulation for electron fit well in the experimental data. The plot indicates that the ion blocking power of $O(10^{-4})$ is achievable with operation provided about -16~V to the gating device, and even higher ion blocking power like $O(10^{-6})$ is possible to reach by adding additional a few volts.

\vspace{2mm}

\section{Resolutions of the full detector}

For verification of the performance that the full detector possesses, such as dE/dx, Z and $r\phi$ resolutions, beam test campaign was conducted at the Test Beam Facility at DESY Hamburg (Germany) with 5~GeV electron beam provided from the DESYII accelerator \cite{desy2}. Electronics for the experiment including a solenoid magnet and a prototype TPC are prepared at the test beam facility, which are commonly used by the LC-TPC collaboration to achieve uniform experimental conditions for all groups to report the performance results of their device. The electronics we used for the experiment are shown and explained in detail in the paper \cite{Ralf}.   


\begin{figure}[htbp]
\begin{center}
	\includegraphics[width=90mm]{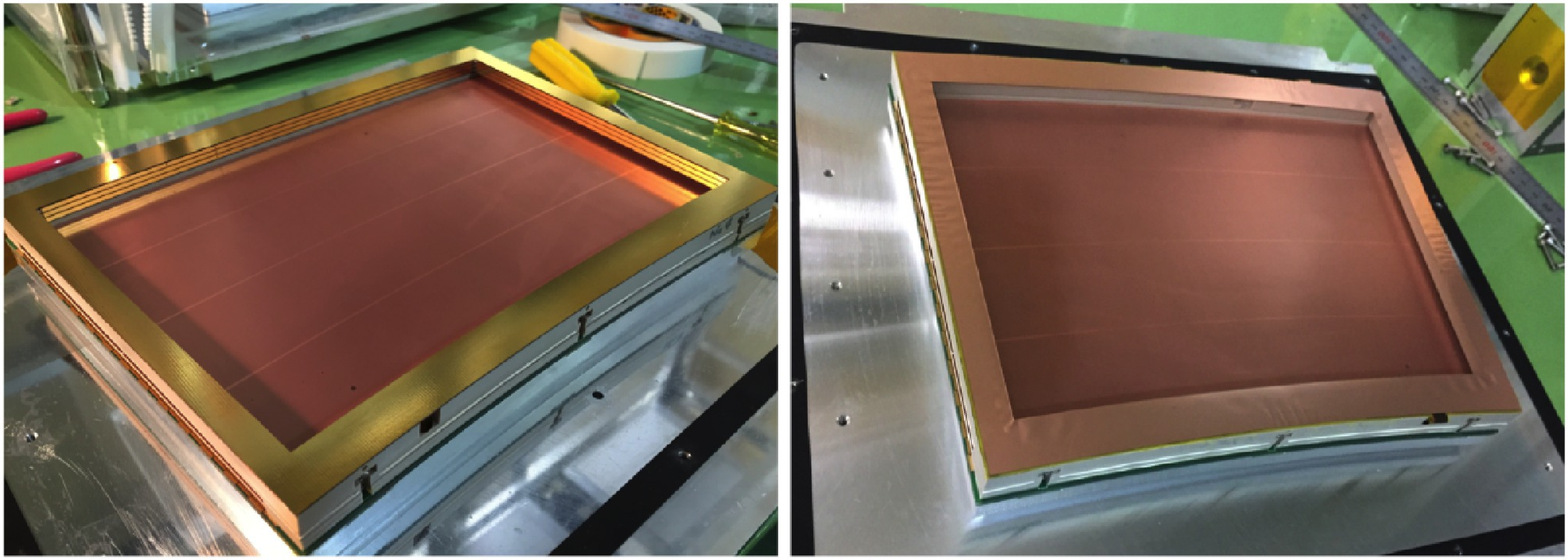}
\caption{Pictures show the modules used for the beam test campaign. The left one is the module with the field shaper, and the right one is the module which the gating device is mounted on.}
\label{fig:fig6_0}
\end{center}
	\vspace{2mm}
\begin{center}
	\includegraphics[width=90mm]{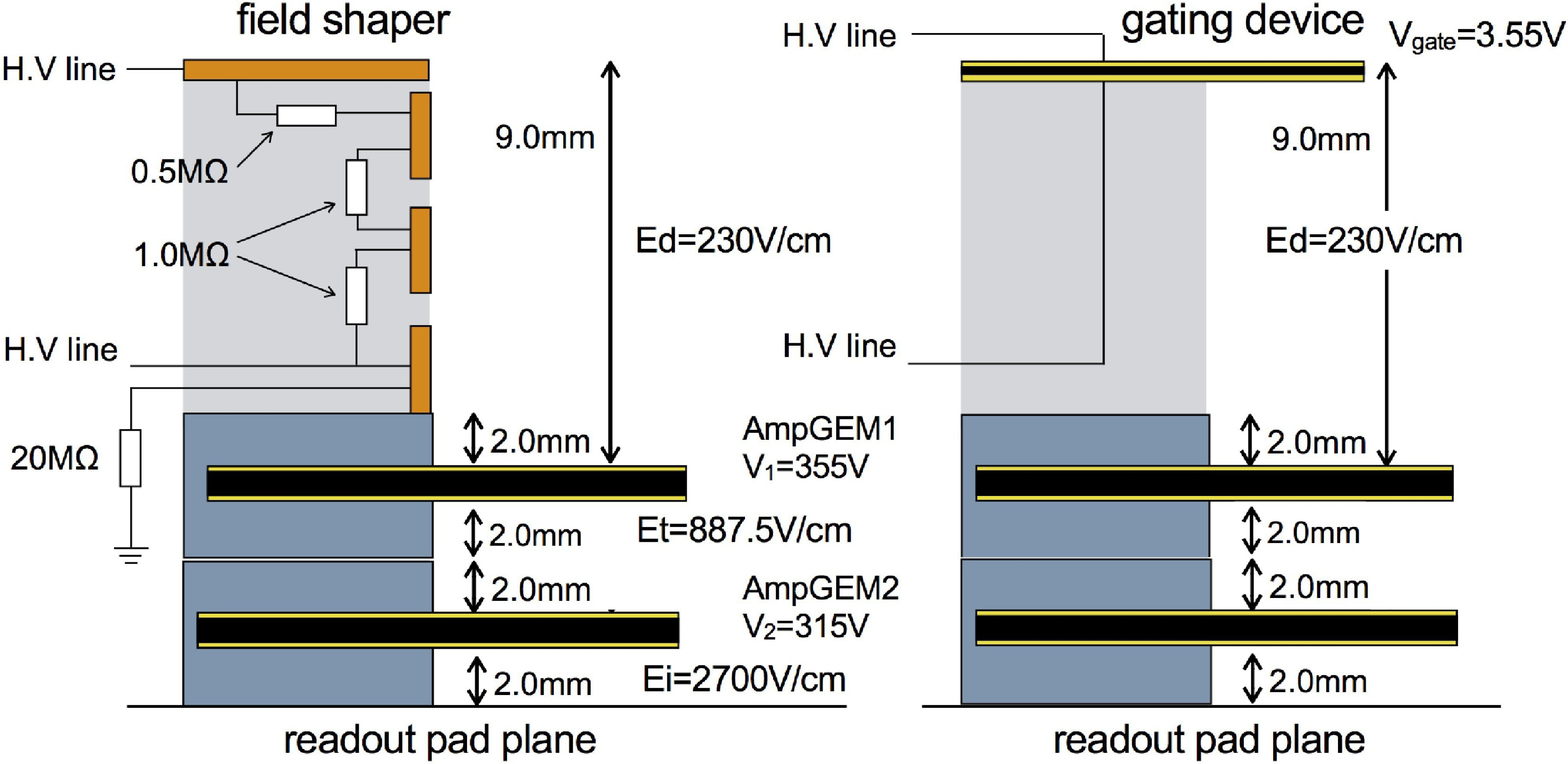}
\caption{A schematic view shows the double GEM configuration with additional upper devices which are the field shaper (left) and the gating device (right).}
\label{fig:fig6_1}
\end{center}
\end{figure}

\subsection{Detector setup}

To confirm the performance of the gating device, two upper devices are prepared. One is, of course, the gating device, and another one is the so-called `field shaper' module without the gate which is used for shaping the electric field between the upper surface of the module and the upper amplification GEM. For the amplification the 100~${\rm \mu m}$ thickness GEM with the diameter of 70 ${\rm \mu m}$ was used with the double stack configuration. The detector configuration for both of the modules can be seen in \Figref{fig:fig6_1}. The measurement for comparing the performance was done by just replacing the upper devices, and keeping the configuration of the amplification GEMs. The modules were operated with the gas gain of around 8,000 during the measurement. The gas gain, although several explanations are given in the later section, was evaluated from a charge distribution of the reconstructed hit object, the setting of the electronics \cite{ALTRO}: 12~mv/fc (PCA16) and 1.17~mV/ADC (ALTRO), the number of primary ionized electrons produced by 5~GeV electron which is 1.4 times larger than MIP and the length of one readout pad (0.5~mm).         


\subsection{Event reconstruction and track selection}

All analysis for beam test data have been done using the MarlinTPC \cite{M-TPC} software package which is common software to create results of LC-TPC collaboration.

Charge clouds after the amplification process in the GEM stack reach the readout pad plane, where the charges are acquired and digitized by the ALTRO chip \cite{ALTRO} with a sampling rate of 20~MHz. To construct a hit object which are composed of several cluster objects in a pad-row and used for track finding and fitting, the clusters have to be distinguish from raw pulse distributions of each pad by giving several parameters: 
a threshold for distinguishing a pulse (5 ADC counts are set), pre-pulse and post-pulse information after the threshold in order to recover left-over parts of the pulse (2 and 5 time bins are set), a time width of the pulse itself (4 time bins are set). The obtained cluster objects in each pad-row are merged into the hit objects whose informations such as position and timing are also given by merging the information of each cluster based on the charge-weighted position (time) method; $x^{clus} = ( \sum_i Q_i \cdot x_i)/ \sum_i Q_i $, where $x$ and $Q$ are position and charge of the pulse, and $i$ denotes the number of pads in the corresponding pad-row. The hit objects of each pad-row are put in as input observables for later track finding and fitting. The track fitting is executed based on TrackMakingKalmanFilterProcessor \cite{Bo} implemented in the MarlinTPC.

In the measurement 20k events were acquired for each data point corresponding to each drift length. Events containing 2 reconstructed tracks or more are excluded from the analysis because such tracks could have scattering with the wall of, for instance, the magnet and deposit energy. After this selection the remaining 1 track events is 16.5k. Additionally azimuthal $\phi$ angle cut is also applied to extract straight line tracks for the performance evaluation, where $0.05^\circ < \phi_0 < 3.39^\circ$ is applied in the analysis coordinate system. The number of events used for the evaluation is 12.6k which is 76~\% of all one track events.

\subsection{Charge distribution}

\begin{figure}[htbp]
\begin{center}
	\includegraphics[width=80mm]{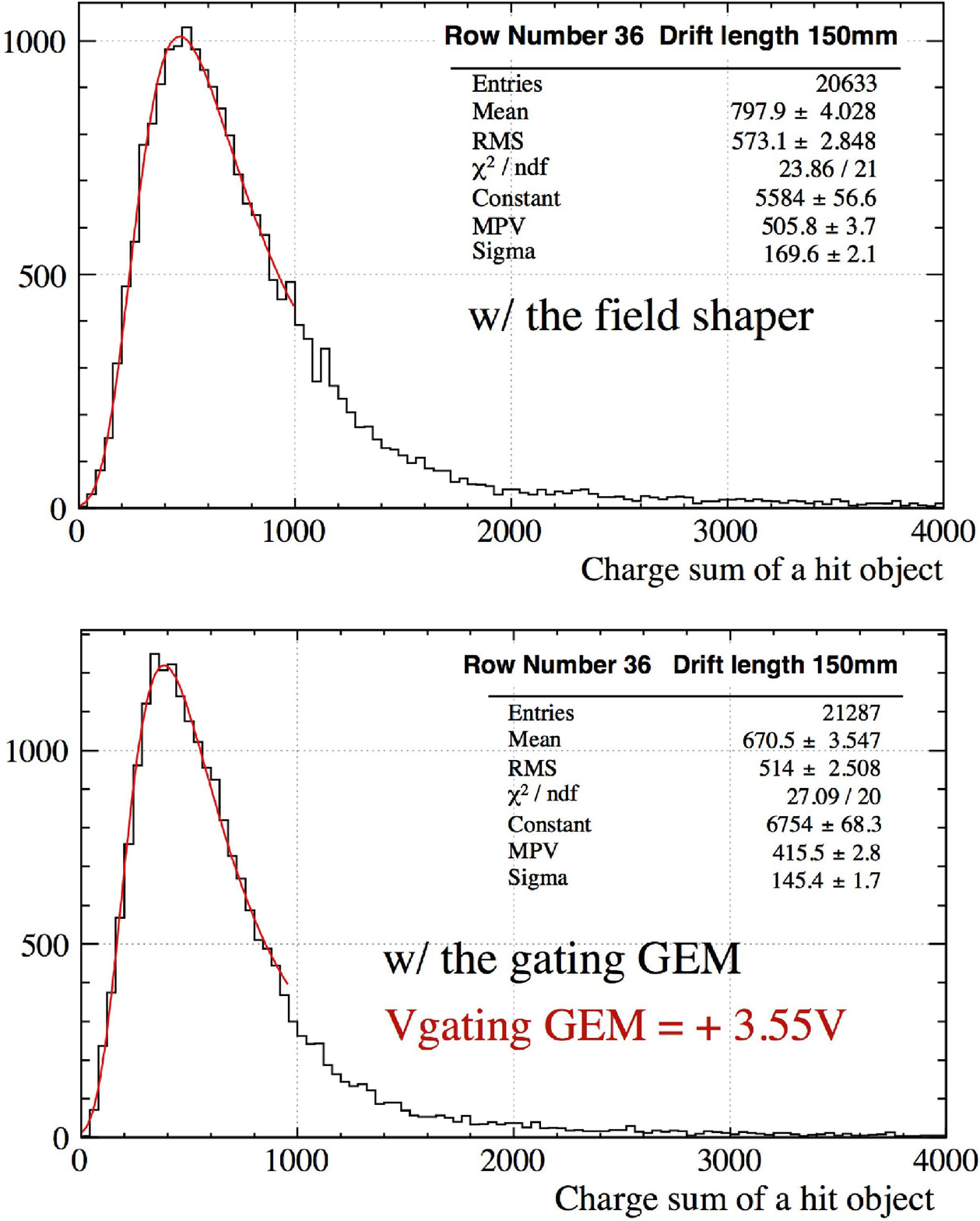}
\caption{The charge distributions of the hit object measured with the module with the shaper (upper) and the gate (lower). Both of the distributions are fitted with the landau function to extract MPVs.}
\label{fig:fig7}
\end{center}
\end{figure}

\Figref{fig:fig7} show charge distributions measured by the module with the shaper and the module with the gate which +3.55~V is applied to. It can be seen that the Most Probable Values (MPV) of both distributions are different, and the ratio of two MPVs is 82.2~$\pm$~0.8~\% which is the consistent value with the electron transmission rate evaluated in the operation with 3--5~V.

\subsection{dE/dx resolution}

\begin{figure}[htbp]
\begin{center}
	\includegraphics[width=74mm,height=64mm]{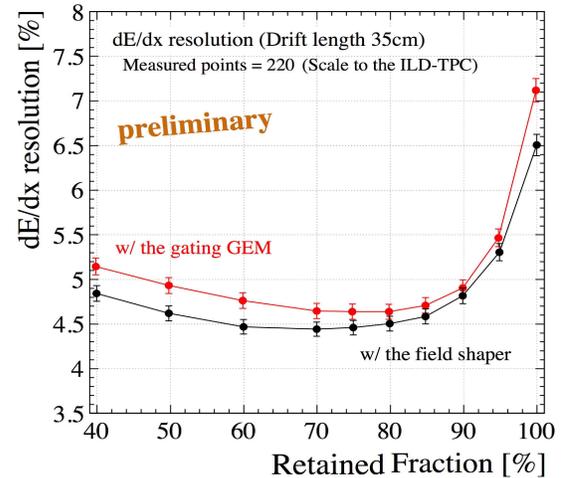}~~~
\caption{The dE/dx distribution as a function of the retained fraction. The results are artificially scaled to the actual size of the ILD-TPC which has 220 sampling points. The black and red color correspond to the module with the shaper and the gate, respectively.}
\label{fig:fig8}
\end{center}
\end{figure}

After track fitting every detector performances are given by using track associated hits and track itself. The dE/dx resolution can be evaluated by using the track associated hits. Because charge distributions have a tail deriving from large energy loss due to ejection of delta ray on the ionization process, the traditional truncated mean method is employed to remove such contributions having larger signal as fractions, where the mean of the energy loss is calculated from large number of sampling points (hit objects). \Figref{fig:fig8} shows the dE/dx resolution as a function of the fraction. The result is scaled to an actual size of the ILD-TPC which has 220 sampling points, by increasing the sampling number artificially. It can be seen that the dE/dx resolution of less than 5~\% is achievable by setting the fraction around 70--80~\%.

\subsection{Z resolution}

\begin{figure}[htbp]
\begin{center}
	\includegraphics[width=85mm]{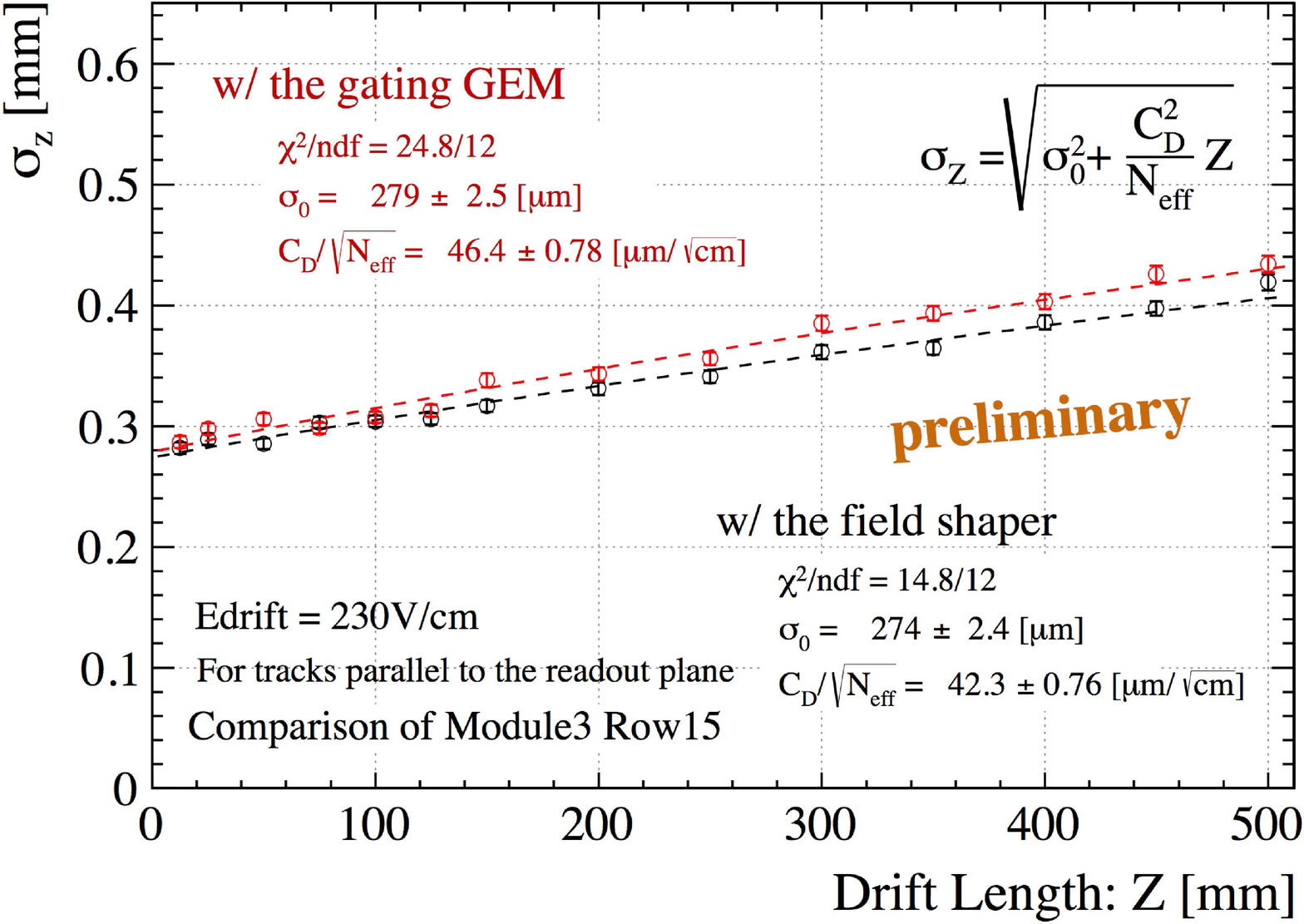}
\caption{The plot shows the Z resolution for tracks parallel to the readout plane as a function of the drift length on both of the modules. The black and red color correspond to the module with the shaper and the gate.}
\label{fig:fig9}
\end{center}
\end{figure}

Determination of arrival time of the pulse is essential for the evaluation of the Z resolution. Several time estimators \cite{Zresol} have been considered and applied to evaluate it. The best way to get the better Z resolution currently is to use a inflection point which is determined by focusing on rising edge of the pulse distribution. The inflection point is calculated based on the derivative-weighted time estimation which is defined as $t^{clus} = ( \sum_i dQ/dt_i \cdot t_i^{bin})/ \sum_i dQ/dt_i $, where $t$ and $Q$ show time and charge on $i$-th bin of the pulse on certain pad.  
The Z resolution can be given by fitting a residual distribution that shows difference of the distance between time of the hit object and the track position, and considering the geometric mean \cite{rphiReso}. \Figref{fig:fig9} shows the Z resolution for the both modules focusing on one arbitrary pad-row. To extract $N_{eff}$ from the data, fitting is performed by using \Equref{eq1} over the drift length. The averaged values of $N_{eff}$ evaluated from center 20 rows are 27.0~$\pm$~0.9 for the module with the shaper, and 22.3~$ \pm$~0.7 for the module with the gating device, where the longitudinal diffusion constant is assumed to be 220~${\rm \mu m/\sqrt{cm}}$ which is calculated with Magboltz under the electric field of 230~V/cm. Because spread of the charge along the time direction is dominated by the shaping time of the electronics, the evaluation of the longitudinal diffusion is not easy.  

Because metal poles connected to the ground are embedded on the current design of the module in order to support the upper structures such as the shaper and the gate, and there are no electrodes on the current gate device for shaping the electric field of the drift volume under the gating device as shown in \Figref{fig:fig6_1}, the outer 4 rows and inner 4 rows are excluded from the evaluation. The ration of $N_{eff}$ between both modules is 82.7~$\pm$~3.8~\% which is the similar value with the other evaluations.

Since the measurement was conducted with 5~GeV electron and the requirements for the ILD-TPC given in the first section are for the MIP, it is necessary to scale the result to the MIP level because degree of energy loss are different. Relative difference of the electron energy loss in the T2K gas was calculated by using HEED simulation \cite{heed}, and the result that the energy loss of the 5~GeV electron is larger with the factor 1.4 compared with the MIP is given. After scaling 5~GeV data to the MIP level, the expected Z resolution for full 2.2~m drift length is 865~${\rm \mu m}$ which satisfies the requirement of the Z resolution for the ILD-TPC. And the averaged $\sigma_Z(0)$ corresponding to 0 drift is 270~${\rm \mu m}$.


\subsection{$r\phi$ resolution}

\begin{figure}[htbp]
\begin{center}
	\includegraphics[width=85mm]{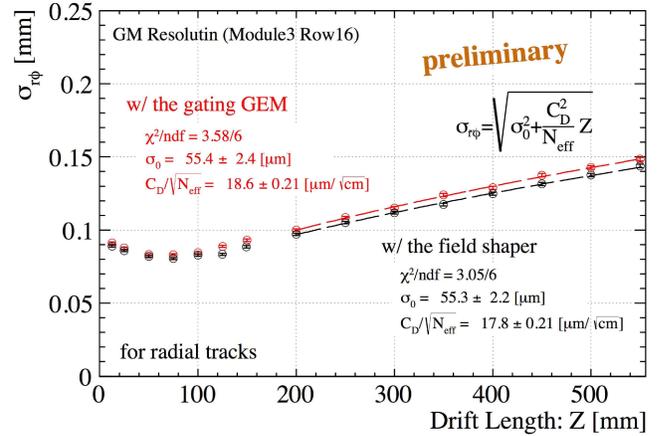}~~~~
\caption{The plot shows the $r\phi$ resolution for radial stiff tracks as a function of the drift length for both of the modules. The fitting for extracting $N_{eff}$ is performed above the region of more than 200~mm to avoid bias because of the finite pad width. The black and red colors correspond to the modules with the shaper and the gate.}
\label{fig:fig10}
\end{center}
\end{figure}

The $r\phi$ resolution is also given by considering the geometric means and fitting the residual distribution that shows difference of the distance between the position of the hit object and the track position. \Figref{fig:fig10} gives the result of the $r\phi$ resolution for both of the modules as a function of the drift length. Similarly the results are fitted with \Equref{eq1} to extract $N_{eff}$ and extrapolate the results to the case that a higher magnetic is assumed. The fitting is performed for drift distances greater than 200~mm to avoid the influence of the bias of the finite pad width which clearly appears for the short drift distance. 

The averaged values of $N_{eff}$ evaluated from center 20 rows are 28.5~$\pm$~0.4 for the module with the shaper, and 22.8~$\pm$~0.3 for the module with the gating device, where the transverse diffusion constants used for the calculation of $N_{eff}$ are also evaluated from the measurement of the pad response width \cite{rphiReso} by fitting with a Gaussian function, which are 95.57~$\pm$~0.24~${\rm \mu m/\sqrt{cm}}$ and 91.95~$\pm$~0.25~${\rm \mu m/\sqrt{cm}}$, respectively. The difference of the transverse diffusion constants measured by both of the modules is presumably derived from intensity of the beam and space charge density resulting from it. However the beam intensity was not unfortunately stored as record during the beam test. The ration of $N_{eff}$ between both modules is 80.1~$\pm$~1.2~\% which is also the consistent observation with the other measurements.

\begin{figure}[htbp]
\begin{center}
	\includegraphics[width=86mm]{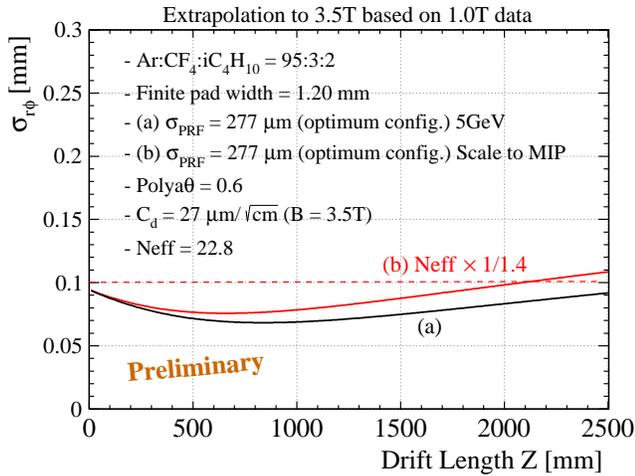}
\caption{The plot shows the $r\phi$ resolution for radial stiff tracks as a function of the drift length. The results are extrapolated to 3.5~T based of the result of 1.0~T. The black line is the direct result using 5~GeV electron, and the red line is scaled to the MIP level by multiplying the factor 1/1.4.}
\label{fig:fig11}
\end{center}
\end{figure}

When a higher magnetic field is added, the transverse diffusion constant varies. Under the ILD-TPC environment the magnetic field of 3.5~T (or 4.0~T) will be added. To predict the performance of the full detector module under such condition, extrapolation is performed, where the transverse diffusion constant under 3.5~T is assumed to be 27~${\rm \mu m/\sqrt{cm}}$ which is taken from a plot showing diffusion constant vs electric filed \cite{T2Kprop}. Since the data has to be also scaled to the MIP level in order to predict the performance for the MIP, the factor 1/1.4 is multiplied $N_{eff}$. The extrapolated results are given in \Figref{fig:fig11}. The corresponding value of $\sigma_{r\phi}$ at full 2.2~m drift length is 103~${\rm \mu m}$ that is almost 100~${\rm \mu m}$.       

\section{Summary}
 
For the future ILC project the MPGD-based TPC is the candidate of the ILD, where the ion back flow is the critical and challenging issue for the realization of the ILD-TPC. To accomplish the required performance of the ILD-TPC and maintain the intrinsic performance of the detector, the large aperture GEM-like gating device was developed. Toward the preparation of the full detector module on which the gating device is mounted for suppressing the ions generated through the amplification, the gating device expanded the sensitive area has been also manufactured. In order to confirm the performance of this gating device, the bench tests on the electron transmission and the ion blocking measurements were conducted. As a result it turned out that the electron transmission rate of 80~\% and the ion blocking power of $O(10^{-4})$ are possible to achieve when the gating device is operated with the lower voltage and the case that -15~V is applied. 

Additionally the test beam campaign was also carried out for the evaluation of the resolutions that the full detector module composed of the double GEM structure and the gating device can provide, and for confirmation of the behavior of the gating device. It was confirmed that the actual size of the ILD-TPC having 220 sampling points can provide the sufficient dE/dx resolution of less than 5~\%, which was evaluated by increasing the sampling points of the beam data artificially. After scaling the 5~GeV results of the Z resolution to the MIP level, it was also confirmed that the Z resolution at 2.2~m full drift length with the full detector module can achieve 865~${\rm \mu m}$. And likewise it was observed that the $r\phi$ resolution which is extrapolated to the 3.5~T condition by using $N_{eff}$ evaluated at 1.0~T and scaled to the MIP level can achieve almost 100 ${\rm \mu m}$. Therefor, the required momentum resolution, which is the most important performance as the tracker, for the ILD-TPC can be attained. 

Each evaluation on the electron transmission under the 1~T magnetic field: the direct measurement of the electron transmission, the comparison of the MPVs of the charge distributions and $N_{eff}$ extracted from the results of the Z resolution and the $r\phi$ resolution, give the consistent values with the geometrical aperture of the gating GEM, which are the results we expected and a proof that the measurements are under our control. The experimental data measured with the full detector module composed of the double GEM structure and the gating device indicate that the detector configured with the double GEM and the gating device satisfies the all requirements of the performance for the ILD-TPC.  
 
\vspace{3mm}
 
\hfill November 9, 2017

\appendices


\vspace{2mm}

\section*{Acknowledgment}
The author would like to thank company staff members of Fujikura Ltd. for their continuous engineering support and production efforts for the gating device. The author would also like to acknowledge the meaningful discussions and comments from the LC-TPC collaboration members. This work was supported by the grant-in-aid for promoted research No.23000002 and No.16H02173 of the Japan Society of Promotion of Science (JSPS).

The measurements leading to these results have been performed at the Test Beam Facility at DESY Hamburg (Germany), a member of the Helmholtz Association (HGF).


\ifCLASSOPTIONcaptionsoff
  \newpage
\fi

\end{document}